\documentclass[a4paper, 12pt]{article}

\usepackage{amsmath}
\usepackage{amssymb}

\def\e{\mathop{\rm \varepsilon}\nolimits}

\newcommand{\Id}{{\rm Id}}
\newcommand{\Pol}{{\rm Pol}}
\newcommand{\su}{{\rm su}}
\newcommand{\SU}{{\rm SU}}
\newcommand{\Tra}{{\rm Tr}}
\newcommand{\Vect}{{\rm Vect}}

\newcommand{\Pt}{{\rm Parot}}
\newcommand{\BiPt}{{\rm BiParot}}
\newcommand{\Gea}{{\rm Gear}}
\newcommand{\SO}{{\rm SO}}
\newcommand{\Span}{{\rm Span}}

\title{Generalized exchange operators for a system of spin-1 particles}

\author{Charlie Jeudy \&  Michel Rouleux}

\begin{document}

\maketitle

\centerline{Aix-Marseille Univ, Universit\'e de Toulon, CNRS, CPT, Marseille, France}

\centerline{charlie.jeudy@etu.univ-amu.fr, rouleux@univ-tln.fr}

\begin{abstract}
The irreps $(\SU(2),{\cal H},U)$ of SU(2) of dimension $(2S+1)^N$,
i.e. operators acting on the space ${\cal H}={\cal H}_N={\bf C}^{(2S+1)^N}$ of $N$ identical particles with spin $S$,  
are described by Clebsch-Gordan decomposition into inequivalent irreps.
In the special case $S=1/2$, Dirac \cite{Dir1}  discovered that 
there is another rep given by $({\cal S}(N),{\cal H},V)$ where ${\cal S}(N)$ is the permutation group,
Thus, the standard ``linear'' Hamiltonian, or Heisenberg interaction Hamiltonian  $H_0=\sum_{1\leq i\leq N}\vec S_i\cdot\vec S_j$, 
where $\vec \sigma_i=2\vec S_i$ is the vector of Pauli matrices, can be 
interpreted as the sum of the ``Exchange Operators'' $P_{ij}$ between particles $i$ and $j$.
Schr\"odinger \cite{Sch} generalized to higher spin numbers $S$ the Exchange Operator $P_{ij}=P_S(\vec S_i\cdot \vec S_j)$ as a polynomial of degree $2S$
in $\vec S_i\cdot \vec S_j$. This we call the $P$-representation. There is another rep induced by
the one particle permutation of states operators $\widetilde Q_\alpha$, which we call the $Q$-rep.
Our main purpose is to write some physical Hamiltonians for a few particles in the $P$- or $Q$-rep and compute their spectrum.
The simplest case where there are as many particles as available states for the spin operator along the $z$-axis,
i.e. $N=2S+1=3$, see Weyl \cite{Wey} or Hamermesh \cite{Ham}. 
Finally, we consider the relationship between permutations and rotation invariance when $S=1/2$ and $S=1$. 
\end{abstract}

\section{\bf Rotational degeneracy and multiplets~: a review}

The situation is best understood for spin 1/2, which fortunately encompasses a great variety of particles. However spin 1 particles,
such as the gauge $W_\pm$ bosons, not to speak of atoms \cite{Dal}, \cite{Ani} play also an important role.

Recall that the irreducible representations (irreps) $(\SU(2),{\cal H},U)$ of SU(2) of dimension $(2S+1)^N$,
i.e. operators acting on the space ${\cal H}={\bf C}^{(2S+1)^N}$ of $N$ identical particles with spin $S$,  
are described by Clebsch-Gordan decomposition into inequivalent irreps defined inductively 
by $D\otimes D'=\bigoplus_\rho m_\rho D^{(\rho)}$.
The group SU(2) enters as a rotational symmetry group of the system.

Characters of these irreps are well known \cite{Jin}.

The easiest case is met for spin $S=1/2$, where SU(2) acts directly 
on the $N$ particles wave function $u=u_1\otimes\cdots\otimes u_N$ by $A\mapsto U(A): (u\mapsto Au_1\otimes\cdots\otimes Au_N)$.
This gives the unitary rep $(\SU(2), {\cal H}, U)$.

The permutation group ${\cal S}(N)$, acting as $\tau\mapsto V(\tau):(u\mapsto(u_{\tau(1)}\otimes\cdots\otimes u_{\tau(N)})$
enters as another symmetry group, because the particles are identical.
This gives the rep $({\cal S}(N), {\cal H}, V)$. Correspondence between reps of SU(2) and ${\cal S}(N)$ is called Schur-Weyl duality.

We have $V(\tau)U(A)=U(A)V(\tau)$. The reps $({\cal S}(N), {\cal H}, V)$ and $(\SU(2), {\cal H}, U)$ are mapped onto each other by Schur-Weyl duality and
are equal up to a multiple of identity, see \cite{Jost} Satz 2.47. 

This was discovered by Dirac \cite{Dir1}, \cite{Dir2}, independently of the theory of group representations
set up a bit earlier by Weyl \cite{Wey}, and expressed in terms of the so-called
{\it exchange operator} between spin-1/2 particles 1 and 2 defined as
\begin{equation}\label{0}
  P(12)=\frac{1}{2}\big(1+\vec \sigma_1\cdot\vec \sigma_2\big)=\begin{pmatrix}1&0&0&0\\ 0&0&1&0\\ 0&1&0&0\\ 0&0&0&1\end{pmatrix}
  \end{equation}
where $\vec\sigma_1\cdot\vec\sigma_2=\sigma^x_1\otimes\sigma^x_2+\sigma^y_1\otimes\sigma^y_2+\sigma^z_1\otimes\sigma^z_2$ (Pauli matrices),
and $\vec \sigma_i=2\vec S_i$ is the vector $(\sigma^x_i,\sigma^y_i,\sigma^z_1)$ is the vector of Pauli matrices at site $i$. 

We recognize the characteristic action of Exchange Operators given by the rep $P(12)$,
$$P(12)|\uparrow\uparrow\rangle=|\uparrow\uparrow\rangle, \ P(12)|\uparrow\downarrow\rangle=|\downarrow\uparrow\rangle,
\ P(12)|\downarrow\uparrow\rangle=\uparrow\downarrow\rangle, \ P_{12}|\downarrow\downarrow\rangle=|\downarrow\downarrow\rangle$$
so that $(P(12))^2=1$.

Following \cite{Jin}, we call {\it class operators} of a finite group those elements of the corresponding group algebra that
one obtained by summing all the group elements that belong to the same conjugacy class -they are also known as ``conjugacy class sums''. 
Thus the standard (linear) interaction Hamiltonian for $N$ identical spin 1/2 particles
(i.e. Heisenberg model with complete graph interaction, including ``self interaction'', and suitably normalized),
is the class operator associated with transpositions $(ij)\in {\cal S}(N)$:
\begin{equation}\label{1}
  H_0(1/2,N)=\sum_{1\leq i<j\leq N}P(ij)
\end{equation}
Here we have identified $\tau\in {\cal S}(N)$ with $V(\tau)$.   
Dirac also allowed for orbital momentum. In total space $L^2({\bf R}^{3N})\otimes{\cal H}$,
the wave function is invariant under simultaneous exchange of position and spin variables,
so the exchange operator is relative to the spin variables alone. 
For simplicity we consider throughout Hamiltonians depending on spin variables only.

More generally, let $\vec S$ be the vector of spin-$S$ matrices, generators of the spin representation of su(2) of dimension $2S+1$.
For $S=1/2$, $\vec S=\frac{1}{2}\vec\sigma$ are $1/2$ times Pauli matrices, for $S=1$,
$S^x,S^y,S^z$ are linear combinations of Gell-Mann matrices. 
Thus we define the standard linear interaction Hamiltonian for $N$ identical spin $S$ particles as
\begin{equation}\label{2}
  H_0(S,N)=C(N)+2\sum_{1\leq i<j\leq N}\vec S_i\cdot \vec S_j
  \end{equation}
where the multiple of identity $C(N)$
is a sum of self-interaction term $\vec S_i^2 = \vec S_i \cdot \vec S_i$  as in (\ref{1}).

The spectrum of $H_0(S,N)$ is computed according to the following ``Multiplet Method'' in a suitable CSCO (so-called {\it un-coupled}
or {\it coupled basis}), which
makes use of Clebsch-Gordan decomposition (\cite{CohTan}, \cite{Jin}, \cite{Ros}). Recall that the coupled basis for $N$ spin particles 
is given by $b=\big((\vec S_i^2)_{1\leq i\leq N}, J^z, \vec J^2\big)$, where $\vec J=\sum_{i=1}^N\vec S_i=(J^x,J^y,J^z)$,
and $\vec J^2=(\sum_{i=1}^N\vec S_i)^2$. The quantum numbers are $(S,M)$, $S$ runs over half-integers  and
\begin{equation}\label{3}
  \begin{aligned}
  &J^z|M,S,(S_i)_{1\leq i\leq N}\rangle=M|M,S,(S_i)_{1\leq i\leq N}\rangle, \ -S\leq M\leq S\\
    &\vec J^2|M,S,(S_i)_{1\leq i\leq N}\rangle=S(S+1)|M,S,(S_i)_{1\leq i\leq N}\rangle
    \end{aligned}
\end{equation}
The correlation between particles arises when
computing $\vec J^2-N \vec S^2=2\sum_{i<j} \vec S_i\cdot \vec S_j$, so that we can directly deduce the spectrum of $H_0$ from this of $\vec J^2$.

For $N=2$ electrons of spin 1/2, $\vec S^2=\vec \sigma^2/4=3/4$, and composition of angular momentum shows that $\vec J^2$
has eigenvalues $s\in\{1/2+1/2=1,1/2-1/2=0\}$, so that $\vec S_1\cdot \vec S_2=(\vec J^2-3/2)/2$,
has eigenvalues $\lambda$ with multiplicity $[\cdot]$:
$\{\lambda=-3/4 [1], \lambda=1/4 [3] \}$. 
For $N=3$, $\vec J^2$
has eigenvalues $s\in\{1/2+1/2+1/2=3/2,1/2+1/2-1/2=1/2\}$, so that $\vec S_1\cdot \vec S_2$
has eigenvalues $\lambda$ with multiplicity $[\cdot]$:
$\{\lambda=-3/4 [4], \lambda=3/4 [4] \}$ and the {\it sum rule} gives again
$1/2\otimes 1/2\otimes 1/2= 1/2\oplus 1/2\oplus 3/2$.

For $S=1$, $N=3$ we can split off ${\bf C}^{27}$ into eigenspaces of $H_0(1,3)$, their dimension is given by the sum rule:
$$1\otimes 1\otimes 1 = (0 \oplus 1 \oplus 2) \otimes 1 = 1 \oplus 0 \oplus 1 \oplus 2 \oplus 1 \oplus 2 \oplus 3$$
namely 1 scalar (multiplicity 1), 3 spins 1 (multiplicity 9), 2 spin 2 (multiplicity
10) and 1 spin 3 (multiplicity 7). For larger $(S,N)$ the Multiplet method can be implemented in many ways (Young diagrams, graphs,\dots)

However, for $S\geq1$, $H_0(S,N)$ can no longer be interpreted {\it a priori} as a class operator
for the action of ${\cal S}(N)$. So we need another algebraic framework, to possibly reinterprete this Hamiltonian,
as well as other rotation invariant (physical) ones.

\section{\bf Exchange operators, $P$-rep and related Hamiltonians for higher order spin systems}

Dirac exchange operators were generalized by Schr\"odinger \cite{Sch},  for higher spin systems of $N$
particles of spin $S$, in the form $P_S(\vec S_i\cdot \vec S_j)$ where $\vec S$ is the vector of spin-$S$ matrices,
$\vec S=\frac{1}{2}\vec\sigma$ for $S=1/2$, and $P_S(x)$ a polynomial of degree $2S$, given by the Ansatz, see also H.A. Brown \cite{Bro}:
\begin{equation}\label{6}
  P_S(x)=(-1)^{2S}\bigl(1+\sum_{p=1}^{2S}{(-2)^p\over(p!)^2}\prod_{q=1}^p(x-x_q)\bigr)=\sum_{n=0}^{2S}A_n x^n
\end{equation}
where $x_q={1\over2}q(q-1)-S(S+1)$, $q=1,2,\cdots,2S+1$, are the eigenvalues of $x$, and $P_S(x_q)=(-1)^{2S+q-1}$.
We can then form the multi-particle (non-linear) Hamiltonian (Schr\"odinger Hamiltonian)
  \begin{equation}\label{7}
    H_S(S,N)=\sum_{\langle i,j\rangle} P_S(\vec S_i\cdot \vec S_j)
  \end{equation}
  Since $H_S(S,N)$ depends only on the scalar products $\vec S_i\cdot \vec S_j$, it trivially commutes with the total rotations of ${\bf R}^3$
  (moving particles all together).
  
  Such Hamiltonians generalizing Heisenberg model,
  are commonly discussed in Statistiscal Mechanics. See also \cite{Josph} for non linear Ising models, using instead polynomials
  $Q_S(S^z_i)$ which are constructed along the same pattern as $P_S(\vec S_i\cdot \vec S_j)$.

  By composing exchange operators $P_S(\vec S_i\cdot \vec S_j)$ relative to transpositions, we easily get~:

  \noindent {\bf Proposition 1}: {\it Let ${\cal H}={\bf C}^{(2S+1)^N}$, there exists a unitary rep $({\cal S}(N),{\cal H}, V_S)$
    of dimension $(2S+1)^N$, defined on generators $(ij)$ of ${\cal S}(N)$
    by $V_S:(ij)\mapsto V_S(ij)=P_S(\vec S_i\cdot S_j)$. }\\

  We denote $V_S((12)(23))=\Pol_S(123)$, etc. the corresponding operator valued polynomials. Thus $\Pol_S(123)$ is of degree 3 with respect to
  $\vec S_i\cdot S_j$.  

  The Class-$k$ Operator associated with $({\cal S}(N),{\cal H}, V_S)$ is defined by summing over all $g_k$ elements belonging to the class $k$.
  This gives
    $$C_k=\sum_{\ell=1}^{g_k}\Pol_S(\tau_\ell)^{(k)}$$
  where $\tau_\ell$ ranges over the $g_k$ Young diagrams. Example with $N=3$: $g_0=1$ (identity element)
  $g_1=3$ (transpositions), $g_2=2$ (circular permutations). In particular for $(S,N)=(1,3)$,
  $\Pol_S(123)$ permutes circularly all spins of particles 1,2,3,
    namely $\Pol_S(123)|\uparrow\uparrow\uparrow\rangle=|\uparrow\uparrow\uparrow\rangle$, 
 $\Pol_S(123)|\downarrow\uparrow\uparrow\rangle=|\uparrow\uparrow\downarrow\rangle$, etc. We have $\Pol_S(123)^3=\Id$.

    This we call the $P$-representation. As in the spin-1/2 case (Dirac)
    we can retrieve Schr\"odinger Hamiltonian $H_S(S,N)$ as the $C_1$ class operator in the $P$-rep. Recall that there is some basis
    of ${\bf C}^3$ where the CSOC-I (see \cite{Jin}, p.53) rep matrices $D(C_i)$ of all class operators for ${\cal S}(3)$ are diagonal,
    in particular $D(C_1)$
    has eigenvalues $(-3,0,3)$, see \cite{Jin}, p.56. 

    The spectrum of $H_S(1,3)$ cannot be computed by the Multiplet Method as for the linear Hamiltonian $H_0(1,3)$. 
For 3 particles of spin 1 however we get, with the help of Mathematica
(Symbolic Calculus on tensor products) the following:\\

\noindent {\bf Proposition 2}:
      {\it For $(S,N)=(1,3)$, Schr\"odinger Hamiltonian $C_1=H_S(1,3)$ (sum over transpositions) has the same eigenvalues $\rho\in\{-3,0,3\} $ as
      those constituting the CSCO-I (see \cite{Jin}, p.53) in the group space of ${\cal S}(3)$.
      Consequently, we have the decomposition into irreducible spaces $\mathcal{L}_{\rho}$
      \begin{equation}
        {\bf C}^{27}=8{\cal L}_0\oplus 10{\cal L}_3+{\cal L}_{-3}
      \end{equation}
      where ${\cal L}_0$ has dimension 2, ${\cal L}_3$ and ${\cal L}_{-3}$ have dimension 1.
      In particular, the character in the $P$-representation associated with $C_1$ is $\chi_P(C_1)=\Tra (C_1)=\Tra (H_S(1,3))=27$}.\\
      
      Note that $m_{\rho}$ factors are deduced by dividing multiplicities of each eigenvalues $\rho$ by the dimension of the irreps $D^{(\rho)}$.\\

\section{\bf State permutation group, $Q$- and $O$-reps}

Instead of exchanging particles, we can permute their states. This gives another representation, which was pointed out by
\cite{Wey}, \cite{Ham}.
It is most easily understood when there are as many particles as available states for the spin operator along some given direction,
i.e. $N=2S+1$, see \cite{Jin}, p.69.\\

\noindent {\it 1) The $Q$-rep}\\

\noindent {\bf Definition 4}: {\it We call 
  $\widetilde Q$-representation of ${\cal S}(2S+1)$ the {\it permutation representation} $({\cal S}(2S+1), {\bf C}^{2S+1}, \widetilde Q)$
  defined on tranpositions $(ij)$, i.e.  $\widetilde Q(ij)$ is the $(2S+1)\times(2S+1)$ transposition matrix on the one particle space, that
exchanges the spin state $i$ and $j$.
We call $Q$-representation the product representation $({\cal S}(2S+1), {\cal H}, Q)$,
${\cal H}={\bf C}^{(2S+1)^N}$ the $N$-lift of $\widetilde Q$, defined on transpositions $(ij)$ 
by $Q(ij)=\widetilde Q(ij)\otimes\cdots\otimes\widetilde Q(ij)$ ($N$ times).}\\

When $S=1$, 
$$\widetilde Q(12)=\begin{pmatrix}0&1&0\cr1&0&0\cr 0&0&1\end{pmatrix}, \ \widetilde Q(13)=\begin{pmatrix}0&0&1\cr0&1&0\cr 1&0&0\end{pmatrix},
\ \widetilde Q(23)=\begin{pmatrix}1&0&0\cr0&0&1\cr 0&1&0\end{pmatrix}$$
and the operators $\widetilde Q_{123}=\widetilde Q_{12}\widetilde Q_{23}$ and $\widetilde Q_{132}=\widetilde Q_{12}\widetilde Q_{13}$
are associated with cycles.

Actually the permutation matrices $\widetilde Q(ij)$ span a Lie algebra in the algebra of $3\times3$ matrices,
  which is idempotent of order 3, i.e. an iterated commutator or order 3 and beyond is spanned by
  $\widetilde Q(\alpha)$, $[\widetilde Q(\alpha), \widetilde Q(\beta)]$ and $[\widetilde Q(\alpha),[\widetilde  Q(\beta), \widetilde Q(\gamma)]]$
  only.

The $P$- and $Q$-reps are not equivalent, as shows the case $(S,N)=(1/2,2)$. Namely
the exchange operator $P(12)=P_S(S_1\cdot S_2)$ (in the canonical basis) is given by (\ref{0}),
and its diagonal basis is the coupled basis, while 
the matrix of permutation of states in  $Q$-representation, associated with the bi-cycle (14)(23), is 
  $$Q(12)=\begin{pmatrix}0&0&0&1\\ 0&0&1&0\\ 0&1&0&0\\ 1&0&0&0\end{pmatrix}$$
 The diagonal basis for $Q(12)$ is Bell'state basis
  $$b_1=\{\big(|\uparrow \downarrow\rangle)+|\downarrow \uparrow\rangle\big), {1\over\sqrt2}
  \big(|\uparrow \uparrow\rangle)+|\downarrow \downarrow\rangle\big) \}\oplus\{{1\over\sqrt2}
  \big(|\uparrow \uparrow\rangle)-|\downarrow \downarrow\rangle\big), {1\over\sqrt2}
  \big(|\uparrow \downarrow\rangle)-|\downarrow \uparrow\rangle\big)\}$$
  
  We can check that the linear Hamiltonian is not in the $Q$-rep, namely
  for any $(c_0,c_1)$, $H_0(1/2,2)=P(12)\neq c_01+c_1Q(12)$.

  At least in case of $(S,N)=(1/2,N)$
and $(S,N)=(1,3)$ it is easy to see that the $Q(ij)$ span a Lie algebra $L$ of operators on the total space ${\cal H}$
and that the Cayley table for $[Q(\alpha), Q(\beta)]$
  is identical to this of $[\widetilde Q(\alpha), \widetilde Q(\beta)]$. 

    Moreover since that the CSCO of the $P$- and $Q$-reps are the same in this case (see \cite{Jin}, p.70), we expect that
Schr\"odinger Hamiltonian $P_S$, as well as other Hamiltonians, belongs to $L$.\\

\noindent {\it 2) The $O$-rep}\\

We can also apply the {\it generator lifting} for $\widetilde Q$ operators.
For $\sigma\in{\cal S}(2S+1)$, define for $i=1,\cdots,N$
$$O_i=1^{\otimes(i-1)}\otimes \widetilde Q(\sigma)\otimes 1^{\otimes(N-i)}$$
which span a Lie algebra by acting only on particle $i$.
Thus for two spin-1/2 particles, we get two commuting operators $O_1,O_2$, with $O_1^2=O_2^2=1$, and then
form $O_1O_2$ which belong to its universal enveloping algebra ${\cal L}$ in $Q$-rep.
It turns out that the linear Hamiltonian $H_0(1/2,2)$
belongs to ${\cal L}$, namely $H_0(1/2,2)=\frac{1}{2}(1-O_1-O_2+O_1O_2)$, and $H_0(1/2,3)=\frac{3}{2}(1-O_1O_2O_3)$.
However, given a Hamiltonian $H$, checking the condition $H\in{\cal L}$ requires to solve for many $c_k$,
and the situation of course worsens for higher spin and particle numbers.

This procedure generalizes to $N$ particles of spin $S$, but it is 
difficult to identify the Hamiltonians obtained this way which are rotation invariant.

It is interesting to investigate how the permutation group $\mathcal{S}(N)$ and
$SU(2)$ are related in the {\it Universal Enveloping Algebra} (UEA), see \cite{Ful}.
The motivation comes from the fact that, for any semi-simple Lie group, we can define a quadratic invariant $J^2$
by using the appropriate Killing form so that for any $g\in G, [J^2,g] = 0$
For a physical system
where we define the action of a rep $\rho(G)$, this is associated with a conserved quantity 
under the symmetry transformations given by $G$. In particular $J^2=\vec J^2$ in our case with $G=\SU(2)$. 
These elements, however, do not belong to the Lie algebra itself but rather to the UEA.
We have seen that $J^2 \sim H_0(S,N)$ (Heisenberg Hamiltonian) and $H_S \sim C_1$ (Schr\"odinger Hamiltonian),
So the $C_1$ class operator plays the same role as $J^2$ in determining the irreps
of $\mathcal{S}(N)$ and $\SU(2)$. Therefore, we are interested
in whether we can identify them directly within a specific UEA. However, the challenge
is to construct an algebra where both $J^2$ and $C_1$ exist within one of its subspaces, see Sect.5.

\section{Relationship between permutations and rotational invariance}

We can consider the cartesian product $(\SU(2))^N$ the full rotation (semi-simple) group of the system,
acting independently on each particle, and associated with direct sum $\su(2)_N=\bigoplus ^N(\su(2))$.
A {\it partial rotation},
denoted $U(R_i)$, is an operator that rotates the $i$-th particle only, while leaving the other particles unaffected.
By construction, this forms a normal subgroup of $(SU(2))^N$.
The group of {\it total rotations} (same rotation applied on each particle) is a subgroup of $(\SU(2))^N$. Its Lie algebra
contains the set $(\vec S_i, \vec S_i^2)_{1\leq i\leq N}$ defined in (\ref{3}). 

Our first result is to express $P_S(\vec S_i\cdot \vec S_j)$ in terms of the coupled basis. 

  When $N=2$, for all $S$, $P_S(\vec S_i\cdot \vec S_j)$ is (trivially) diagonal in the coupled basis, or in the basis diagonalizing $\vec J^2$.
  For $N$ particles of spin $S$ we can check by Symbolic Calculus for ``small'' values of $(S,N)$ the following:\\
 
  \noindent {\bf Proposition 5}: {\it For $S=1/2,1$ we have:
  
  1) $P_S(\vec S_i\cdot \vec S_j)$ commutes with all elements of the coupled basis (CSCO-II of $SU(2)$, see \cite{Jin}, p.79:
$$[P_S(\vec S_i\cdot \vec S_j),\vec J^2]=[P_S(\vec S_i\cdot \vec S_j),\vec J^z]=[P_S(\vec S_i\cdot \vec S_j),\vec S_k^2]=0$$

  2) Moreover, $P_S(\vec S_i\cdot \vec S_j)$ commutes with SU(2), more precisely with the (global) rotations
  $U(A): (u_1,\cdots, u_N)\mapsto (Au_1\otimes\cdots\otimes Au_N)$. When $S=1$, $A$ can be identified (locally) with an element of SO(3)).
}

  We complete the coupled basis $b$ to $\overline b=\{(\vec S_i^2)_{1 \leq i \leq N}, \vec{J}^2, \vec{J}^z, \overline{J_z} \}$,
  which constitute the CSCO-III of $SU(2)$ with $\overline{J_z} $
  being one generator of the intrinsic group, see \cite{Jin}, p.79. If
  $[P_S(\vec S_i\cdot \vec S_j),\overline{J_z}]=0$, then we could infer 
  from Dirac theorem on joint spectrum of commuting operators that 
  $P_S(\vec S_i\cdot \vec S_j)=F_{ij}\bigl((\vec S_i^2)_{1\leq i\leq N},\vec J^2,\vec J^z, \overline{J_z} \bigr)$
  where $F_{ij}$ can be chosen as a polynomial (Lagrange interpolation polynomial in $N+2$ variables).
Possibly also we can remove $J^z$ from the variables
due to the invariance of $\vec{S}_i \cdot \vec{S}_j$ under rotation generated by $SU(2)$.\\

Next we investigate the commutation relations between partial rotations and exchange operators.
Together with the global rotations above, we consider 3 types of partial rotations. Let $A\in \SO(3)$ 

1) $\Pt(k): (u_1,\cdots, u_N)\mapsto (u_1\otimes\cdots Au_k\otimes u_{k+1}\otimes\cdots u_N)$ (rotating only particle $k$).

2) $\BiPt(ij): (u_1,\cdots, u_N)\mapsto (u_1\otimes\cdots Au_i\otimes u_{i+1}\otimes\cdots \otimes Au_j\cdots u_N)$
(rotating only particles $i$ and $j$)

3) $\Gea(ij):  (u_1,\cdots, u_N)\mapsto (u_1\otimes\cdots Au_i\otimes u_{i+1}\otimes\cdots \otimes A^{-1}u_j\cdots u_N)$ (rotating only particles
$i$ and $j$ in opposite sense).

We have~:\\

\noindent {\bf Proposition 6}: {\it For $S=1/2,1$ we have:

  1) $[P_S(\vec S_i\cdot \vec S_j), \Pt(k)]=0$ for all $i,j,k$.

  2) $[P_S(\vec S_i\cdot \vec S_j), \BiPt(ij)]=0$ for all $i,j$.}\\

Next we list some Casimir operators (still for small $(S,N)$)~:\\

\noindent {\bf Proposition 7}: {\it For $S=1/2,1$ there holds:

  1) $\Pt(k)$ is a normal subgroup of $G=\SU(2)\times\cdots\SU(2)$.

  2) $\BiPt(ij)$ is a normal subgroup of $\SU(2)\approx\{(A,\cdots,A): A\in \SU(2)\}$.

  3) For all $k$, Casimir operator $\Pt(k)=S_k^2\in b$

  4) For all $(i,j)$, $(S_i+S_j)^2$ is a Casimir operator for $\BiPt(i,j)$, but $(S_i+S_j)^2\notin \Span \{b\}$.

  5) For all $(i,j)$, $(S_i-S_j)^2$ is a Casimir operator for $\Gea(i,j)$, but $(S_i-S_j)^2\notin \Span \{b\}$.
}\\

Now we examine the relationship between the permutation group ${\cal S}(N)$ and some enveloping algebra.

\section{Class operators $C_n$ and universal enveloping algebra $U(\su(2)_N)$}

 We denote by $U(\mathfrak{g})$ the \textit{Universal Enveloping Algebra} UEA of a Lie algebra $\mathfrak{g}$,
 defined as the associative algebra of polynomials in elements of $\mathfrak{g}$, where $\mathfrak{g}$ is the Lie algebra of a semi-simple
 Lie group. Consider here $G=G_1=\SU(2)$ and its Lie algebra $\mathfrak{g}_1=\su(2)$, which is also the Lie algebra of $\SO(3)$. 

 Thanks to Poincar\'e–Birkhoff–Witt theorem, a basis of $U(\mathfrak{su(2)}$
 can be constructed from the infinite set of elements $\{J_x^n J_y^m J_z^\ell\}_{n,m,\ell\in{\bf N}_0}$. 
 
 The UEA of a semi-simple Lie algebra $\mathfrak{g} = \mathfrak{g}_1 \oplus \mathfrak{g}_2$, where $\mathfrak{g}_1$
 and $\mathfrak{g}_2$ are simple, is thus given by the isomorphism
 $U(\mathfrak{g}_1 \oplus \mathfrak{g}_2) = U(\mathfrak{g}_1) \otimes U(\mathfrak{g}_2)$.
 So for  $G=(SU(2))^N$ (direct product), and $\mathfrak{g} = \bigoplus^N \mathfrak{su(2)}$ (direct sum),
 a basis of $U(\mathfrak{su(2)}_N)$ is given by  $\{\prod_{i=1}^N (J_x^n J_y^m J_z^\ell)_i\}_{n,m,\ell \in{\bf N}_0}$.

 We may ask the following question: Can the group space of ${\cal S}(N)$ be mapped to a space of polynomials
 (e.g. the family of Schr\"odinger polynomials indexed by the quantum number $S$), valued in $U(\mathfrak{su(2)}_N)$,
 with the additional property that the class operators (for the action of ${\cal S}(N)$)
 belong to the centralizer of the group of total rotations (identified with SU(2)? In particular, $C_1\equiv J^2$, see Proposition 7.

 These class operators can be represented as exchange operators ($P$-rep), are all invariant under rotations,
and commute with Heisenberg Hamiltonian. They would define quantum numbers characterizing collective property of spin systems,
and we can also study their link with \textit{generalized Casimir operators} of $\mathfrak{su(2)}_N$ 

Note also that there is another one particle irrep $(\SU(2),{\cal H}_{2S},W)$ of SU(2) of dimension $2S+1$, where ${\cal H}_{2S}$
  is the space of homogeneous polynomials of degree $2S$ of 2 complex variables, namely $W(A)p(Z)=p(AZ)$,
  $Z={}^t(x,y)$. Taking tensor products this lifts to the $(2S+1)^N$ dimensional rep
  $$W_N(A)(p_1\otimes\cdots\otimes p_N)(Z_1,\cdots,Z_N)=p_1(AZ_1)\cdots p_N(AZ_N)$$
 We obtain in the same way another rep of ${\cal S}(N)$, permuting the variables $Z_i$.

\section{A sub-representation of Lie algebra $(L,{\cal H}, Q)$ and its dual space }

Recall the $Q$-rep is a representation of the Lie algebra $L$ of (lifted) permutation matrices.
We restrain here ${L}$ to a smaller class $\widehat L$, and examine the spectral properties of the Hamiltonians it contains. We expect
$H_S(S,N)\in{\widehat L}$, as well as other physical Hamiltonians. Commutators in the $Q$-rep will play the role of polynomials in the $P$-rep.
To fix the ideas we assume $(S,N)=(1,3)$.

Consider first the Lie algebra $\widetilde L$ (called the ``small'' or ``parent'' algebra) 
of all operators on ${\bf C}^{3}$ generated by the $\widetilde Q_\alpha$, 
and recall from Sect.3.1 that $\widetilde L$  is idempotent of order 3, and its lift $L$ is spanned by the
$Q_\alpha=\widetilde Q_\alpha\otimes\cdots\otimes \widetilde Q_\alpha$ is such that Cayley Tables for commutators in $L$ coincide
with this in $\widetilde L$.

For $\alpha=(ij)$ or $\alpha=(ijk)$ we choose $\e _\alpha$ according to the rules of a group algebra, i.e. $\e _\alpha=1$ if $\alpha$ is a permutation,
  and $\e _\alpha=-1$ if $\alpha$ is a cycle, and we set
  \begin{equation}\label{15}
\begin{aligned}
  &J_1=\sum_\alpha \e _\alpha^3Q_\alpha=Q_{12}+Q_{13}+Q_{13}-Q_{123}-Q_{132}\\
  &J_2=1, \quad
  J_3=[J_1,[J_1, J_1]']'=\sum'\e _\alpha^3 \e _\beta^3 \e _\gamma^3 [Q_\alpha, [Q_\beta,Q_\gamma]]
\end {aligned}
\end{equation}
where $[J_1, J_1]'=\sum_{\alpha\prec\beta}\e _\alpha^3 \e _\beta^3[Q_\alpha, Q_\beta]$, and $\sum'$
means a sum over some ordered subset $\alpha,\beta,\gamma$, which reduces in the present case to
$(\alpha,\beta,\gamma)=((12),(13),(23))$, and $J_3=2[Q_{12}, [Q_{13},Q_{23}]]$.
Thus $J_1$ is the first class operator $C_1$ of the group algebra. We check that $[J_1,J_3]=0$, so that
  $L_1=\Vect(J_1,J_2,J_3)$ is a commutative, unitary 3-D Lie algebra. In the parent algebra the corresponding 
  class operators take the form $\widetilde J_1=\widetilde J_2=\Id$, and
  $\widetilde J_3=2[\widetilde Q_{12}, [\widetilde Q_{13}, \widetilde Q_{23}]]=2(\widetilde Q_{13}-\widetilde Q_{12})$.
  The spectral values $\lambda$ with multiplicity $[m]$
  of any Hamiltonian $H=c_1J_1+c_3J_3\in L_1$ is readily computed from Symbolic Calculus,
  ($\lambda=-3c_1[4]$,$\lambda=3c_1[5]$, $\lambda=-4\sqrt3c_3[9]$, $\lambda=4\sqrt3c_3[9]$)
  but none of these Hamiltonians (up to a constant) is of the form $H_0(1,3)$ or $H_S(1,3)$ and probably not rotation invariant.
  
To remedy this fact, we introduce the dual space $L_1^*$ (for the canonical Hermitian product $\langle A,B\rangle=\Tra AB^*$)
defined in the following way. In the space of all real symmetric $3\times3$ matrices,
  we construct a dual family $\widetilde q_\beta$ of the symmetric part of the $\widetilde Q_\alpha$'s, consisting of 4 terms
  $\widetilde q_{12},\widetilde q_{13},\widetilde q_{23},\widetilde q_{123}=\widetilde q_{132}$, which are real symmetric matrices.
  It is thus only defined modulo some 2-D subspace $\Vect(\widetilde M_1,\widetilde M_2)$, $\widetilde M_1, \widetilde M_2$ real symmetric.
  Then we consider the lift $q_\alpha$ of the $\widetilde q_\alpha$'s, and
  for $\e =-1$ or 0 the
  dual class operators $k_1=q_{12}+q_{13}+q_{13}+\e (q_{123}+q_{132})$, $k_2=[k_1, k_1]'$ and
  $k_3=[k_1,[k_1, k_1]']'$ as in (\ref{15}). Up to a change of basis $(k_1,k_2,k_3)\mapsto(j_1,j_2,j_3)$ we can fulfill the orthogonality
  conditions $\langle j_\alpha,J_\beta\rangle=\Tra (j_\alpha J_\beta)=\delta_{\alpha,\beta}$, also when modifying the $\widetilde q_\alpha$'s
  by a linear combination of matrices $\widetilde M_1, \widetilde M_2$.
  Thus we get the dual space $L_1^*$ of $L_1$. 
  It remains to show that $L_1^*$ is also a Lie algebra, and that $L_1\oplus L_1^*$ becomes a Lie bi-algebra. We expect $L_1^*$
  to be also commutative, so that $L_1\oplus L_1^*$ identifies with a Euclidean phase-space from which we can retrieve SU(2)
  (or rather SO(3) locally) by integration. 
  Now we can form $H=c_1J_1+c_2J_2+ c_3 J_3+d_1j_1+d_2j_2+d_3j_3$ and try to characterize those  which are rotational invariant.
  Thus (examining the multiplicity of eigenvalues) we strongly expect that Schr\"odinger Hamiltonian) $H_S(1,3)\in \widehat L=L_1+ L_1^*$,
  and that there are other Hamiltonians in $L_1+L_1^*$ whose spectrum is integer valued as well, up to trivial factors.
This reflects the fact
that all irreps of ${\cal S}(N)$ are rational.\\


\noindent {\it Acknowledgements}: We thank Walter Aschbacher, Robert Coquereaux, Janos Polonyi, Laurent Raymond and Herv\'e Ricard for useful discussions.
The first Author acknowledges financial support from QuanTEdu-France during his Master 2 Internship at Aix-Marseille Universit\'e. \\

\end{document}